\documentclass[useAMS,usenatbib]{mn2e}
\usepackage[dvips]{graphicx}
\usepackage{amsmath}
\title[Basic Properties of MHD turbulence]{Basic Properties of MHD turbulence in the Inertial Range}
\author{Andrey Beresnyak$^{1,2}$\\
$^{1}$Los Alamos National Laboratory, Los Alamos, NM, 87545\\
$^{2}$Ruhr-Universit\"at Bochum, 44780 Bochum, Germany.}
\begin{document}
\bibliographystyle{mn2e}

\maketitle

\begin{abstract}
 We revisit the issue of spectral slope of MHD turbulence in the inertial range and
 argue that numerics favor Goldreich-Sridhar -5/3 slope rather than -3/2 slope. We also did
 precision measurements of anisotropy of MHD turbulence and determined the anisotropy
 constant $C_A=0.34$ of Alfv\'enic turbulence. Together with previously measured Kolmogorov
 constant $C_K=4.2$, or 3.3 for purely Alfv\'enic case, it constitutes a full description
 of MHD cascade in terms of spectral quantities, which is of high practical value for astrophysics.     
\end{abstract}

\begin{keywords}
MHD -- turbulence.
\end{keywords}
%\keywords{MHD -- turbulence}
%\pacs{52.65.Kj, 52.30.Cv, 47.27.Jv, 95.30.Qd, 52.35.Ra, 47.27.E-, 52.30.Cv}

\section{Introduction} 
Astrophysical plasmas are often described as ideal MHD fluid -- a perfectly conducting,
inviscid fluid described by MHD equations. The use of continuous fluid approach and absence
of dissipation terms is motivated by the fact that astrophysical scales are typically
humongous compared to the molecular and plasma mean free paths and the microscopic dissipation
scales\footnote{This rule have some notable exceptions, such as molecular gas in the early
Universe, which could be fairly viscous on scales of protogalaxies or
the solar wind which is not collisional enough to fully support compressible modes.}.
MHD can be successfully applied to all phases of the interstellar medium (ISM), large
scales of molecular clouds, intracluster medium (ICM), stellar interiors and stellar outflows, etc.
Reynolds numbers of astrophysical flows are, typically, very high, which makes turbulence
ubiquitous. Initially unmagnetized well-conductive turbulent fluid generates its own
magnetic field which is dynamically important on almost all scales. In spiral galaxies
large-scale dynamo generates ordered fields on the scale of the galactic disc, while
in more symmetric objects, such as galaxy clusters, small-scale dynamo operates and
generates fields on scales up to 20 kpc.

Inertial range of turbulence was introduced by \citet{kolm41} as a range of spatial
scales where driving and dissipation are unimportant and perturbations exist due
to energy transfer from one scale to another.
In the inertial range of MHD turbulence perturbations of both velocity and magnetic
field will be much smaller than the local Alfv\'enic velocity $v_A=B/\sqrt{4\pi\rho}$, due to the turbulence
spectrum being steeper than $k^{-1}$, therefore local mean magnetic
field will strongly affect dynamics in this range \citep{iroshnikov, kraichnan}.
As in the case of hydrodynamics, the study of MHD turbulence began with
weakly compressible and incompressible cases which are directly applicable to
many environments, such as stellar interiors, ICM and hot phases of the ISM.
Later it was realized that many features of incompressible MHD turbulence
are still present even in supersonic dynamics, due to the dominant effect of
Alfv\'enic shearing \citep{cho2003c,BLC05}. It had been pointed out by \citet{GS95} that
strong mean field incompressible turbulence is split into
the cascade of Alfv\'enic mode, described by Reduced MHD or RMHD \citep{kadomtsev1974,strauss1976} and the passive cascade
of slow (pseudo-Alfv\'en) mode. In the strong mean field case it was sufficient
to study only the Alfv\'enic dynamics, as it will determine all statistical
properties of turbulence, such as spectrum or anisotropy. This decoupling
was also observed in numerics. Luckily, being the limit
of very strong mean field, RMHD has a precise two-parametric symmetry similar to the
one in incompressible hydrodynamics (see \S 2), which, under certain conditions,
makes universal cascade with power-law energy spectrum possible.

Interaction of Alfv\'enic perturbations propagating in a strong mean field is unusual
due to a peculiar dispersion relation of Alfv\'enic mode, $\omega=k_\| v_A$, where $k_\|$
is a wavevector parallel to the mean magnetic field. This results in a tendency of MHD
turbulence to create ``perpendicular cascade'', where the flux of energy is preferentially
directed perpendicular to the magnetic field. This tendency enhances the nonlinearity
of the interaction, described by $\xi=\delta v k_\perp/v_A k_\|$, which is the ratio of the mean-field
term to the nonlinear term, and results in development
of essentially strong turbulence. As turbulence
becomes marginally strong, $\xi\sim 1$, the cascading timescales
become close to the dynamical timescales $\tau_{\rm casc}\sim\tau_{\rm dyn}=1/wk_\perp$ and
the perturbation frequency $\omega$ has a lower bound due to an
uncertainty relation $\tau_{\rm casc}\omega>1$ \citep{GS95}. This makes turbulence
being ``stuck'' in the $\xi\sim 1$ regime, which is known as ``critical balance''.
Note, that the above argument is not the only lower bound on $\omega$, with
another bound being due to the {\it directional} uncertainty of the ${\bf v}_A$,
which was discovered in \citet{BL08}. In this paper we will mostly consider so-called
``balanced'' turbulence in which both uncertainties coincide. We refer the reader
to \citet{BL08, BL09a} and references therein for the more general imbalanced case.

Goldreich-Sridhar model is predicting a $k^{-5/3}$ energy spectrum with anisotropy
described as $k_\|\sim k_\perp^{2/3}$. Numerical studies \citet{Cho2000, maron2001} confirmed
steep spectrum and scale-dependent anisotropy, but \citet{maron2001, muller2005} claimed
a shallower than $-5/3$ spectral slope in the strong mean field case, which was close to $-3/2$.
This motivated adjustments to the Goldreich-Sridhar model
\citep{galtier2005,boldyrev2005,gogoberidze2007}. A model with so called ``dynamic
alignment'' \citep{boldyrev2005, boldyrev2006} became popular after the
scale-dependent alignment was discovered in numerical simulations \citep{BL06}. This model
is based on the idea that the alignment between velocity and magnetic perturbations
decreases the strength of the interaction scale-dependently, relying on the alignment
being a power-law function of scale. This would, as they argue, modify the spectral slope
of MHD turbulence from the $-5/3$ Kolmogorov slope to the observed $-3/2$ slope.
It was also claimed that there is a self-consistent turbulent mechanism that produces
such an alignment. In this paper we examine both the alignment and the spectrum.
It turns out that the alignment is not universal and is tied to the driving scale.
Also, spectral resolution study at high numerical resolutions favor $-5/3$ spectral
slope more than $-3/2$ slope. It seems that earlier measurements of the MHD slope
were premature and were not conducted with enough rigor.

In this paper we a) support our earlier claim of -5/3 scaling and the measurement of Kolmogorov constant
of \citet{B11} and b) perform the measurement of anisotropy constant. These two measurements allow us to predict the local properties
of MHD turbulence on any scale $l$, given only the dissipation rate $\epsilon$ and Alfv\'en velocity $v_A$. Such a
straightforward prediction is of great practical use for astrophysics.

\section{Basic Equations}
Ideal MHD equations describe the dynamics of ideally conducting inviscid fluid with magnetic field
and can be written in Heaviside and $c=1$ units as

%\begin{equation}
\begin{align*}
\partial_t \rho+{\bf \nabla\cdot}(\rho{\bf v})&=0,\\
\rho(\partial_t+{\bf v}\cdot\nabla){\bf v})&=-\nabla P+{\bf j\times B},\\
{\bf \nabla\cdot B}&=0,\\
\partial_t{\bf B}&=\nabla\times({\bf v \times B}),%\label{mhd0}%\tag{1}
\end{align*}
%\end{equation}

with current ${\bf j}={\bf \nabla \times B}$ and vorticity ${\bf \omega}={\bf \nabla \times v}$.
This should be supplanted with energy equation and a prescription for pressure $P$.
The incompressible limit assumes that the pressure is so high that the density is constant
and velocity is purely solenoidal (${\bf \nabla \cdot v}=0$). This does not necessarily refer to the ratio of outer scale
kinetic pressure to molecular pressure, but could be interpreted as scale-dependent condition.
Indeed, if we go to the frame of the fluid, local
perturbations of velocity will diminish with scale and will be much smaller than the speed of sound.
In this situation it will be possible to decompose velocity into low-amplitude
sonic waves and essentially incompressible component of ${\bf v}$, as long as we are not in the vicinity
of a shock. The incompressible component, bound by ${\bf \nabla\cdot v}=0$,
will be described by much simpler equations: 

%\begin{equation}
\begin{align*}
\partial_t{\bf v}&=\hat S(-{\bf \omega \times v}+{\bf j \times b}),\\
\partial_t{\bf b}&=\nabla\times({\bf v \times b}),%\label{mhd1}%\tag{2}
\end{align*}
%\end{equation}

where we renormalized magnetic field to velocity units ${\bf b=B}/\rho^{1/2}$ (the absence of $4\pi$
is due to Heaviside units) and used solenoidal projection operator
$\hat S=(1-\nabla\Delta^{-1}\nabla)$ to get rid of pressure.
Finally, in terms of Els\"asser variables ${\bf w^\pm=v\pm b}$ this could
be rewritten as

\begin{equation}
\partial_t{\bf w^\pm}+\hat S ({\bf w^\mp}\cdot\nabla){\bf w^\pm}=0.\label{mhd2}
\end{equation}

This equation resembles incompressible Euler's equation. Indeed, hydrodynamics
is just a limit of $b=0$ in which ${\bf w}^+={\bf w}^-$. This resemblance, however, is misleading,
as the local mean magnetic field could not be excluded by the choice of reference
frame and, as we noted earlier, will strongly affect dynamics on all scales.
We can explicitly introduce local mean field as ${\bf v_A}$, assuming that it is constant,
so that $\delta{\bf w^\pm=w\pm v}_A$:

\begin{equation}
\partial_t{\bf \delta w^\pm}\mp({\bf v_A}\cdot\nabla){\delta \bf w^\pm}
+\hat S ({\delta \bf w^\mp}\cdot\nabla){\delta \bf w^\pm}=0\label{mhd3}.
\end{equation}

In the linear regime of small $\delta w$'s they represent perturbations, propagating
along and against the direction of the magnetic field, with nonlinear term describing
their interaction. As we noted earlier, due to the resonance condition of Alfv\'enic
perturbations they tend to create more perpendicular structure,
making MHD turbulence progressively more anisotropic. This was empirically known
from tokamak experiments and was used in so-called reduced MHD approximation, which neglected
parallel gradients in the nonlinear term \citep{kadomtsev1974,strauss1976}.
Indeed, if we denote $\|$ and $\perp$ as directions parallel and perpendicular to ${\bf v}_A$,
the mean field term $(v_A \nabla_\|) \delta w^\pm$ is much larger than
$(\delta w^\mp_\| \nabla_\|)\delta w^\pm$ and the latter could be ignored in the inertial
range where $\delta w^\pm \ll v_A$. This will result in Equation~\ref{mhd3} being split into

\begin{equation}
\partial_t{\bf \delta w^\pm_\|}\mp({\bf v_A}\cdot\nabla_\|){\delta \bf w^\pm_\|}
+\hat S ({\delta \bf w^\mp_\perp}\cdot\nabla_\perp){\delta \bf w^\pm_\|}=0,\label{mhd4}
\end{equation}
\begin{equation}
\partial_t{\bf \delta w^\pm_\perp}\mp({\bf v_A}\cdot\nabla_\|){\delta \bf w^\pm_\perp}
+\hat S ({\delta \bf w^\mp_\perp}\cdot\nabla_\perp){\delta \bf w^\pm_\perp}=0,\label{mhd5}
\end{equation}

which, physically represent a limit of very strong mean field where ${\bf \delta w^\pm_\|}$
is a slow (pseudo-Alfv\'en) mode and ${\bf \delta w^\pm_\perp}$ is the Alfv\'en mode and
Equation~\ref{mhd4} describes a passive dynamics of slow mode which is sheared by the
Alfv\'en mode, while Equation~\ref{mhd5} describes essentially nonlinear dynamics of the Alfv\'en mode
and is known as reduced MHD. For our purposes, to figure out asymptotic
behavior in the inertial range, it is sufficient to study Alfv\'enic dynamics and slow
mode can be always added later, because it will have the same statistics. 

It turns out that reduced MHD is often applicable beyond incompressible MHD limit, in a highly
collisionless environments, such as tokamaks or the solar wind. This is due to the fact that
Alfv\'en mode is transverse and does not require pressure support. Indeed, Alfv\'enic perturbations
rely on magnetic tension as a restoring force and it is sufficient that charged particles
be tied to magnetic field lines to provide inertia \citep{schekochihin2009}.

A remarkable property of RMHD is that it has a precise two-parametric symmetry with respect to the
anisotropy and the strength of the mean field:
${\bf w} \to {\bf w}A,\ \lambda \to \lambda B,\ t \to t B/A,\ \Lambda \to \Lambda B/A
$, which is similar to hydrodynamic symmetry.
Here $\lambda$ is a perpendicular scale, $\Lambda$ is a parallel scale, $A$
and $B$ are arbitrary parameters of the transformation. It is due to this precise symmetry and the absence
of any designated scale, that we can hypothesize universal regime,
similar to hydrodynamic cascade of \citet{kolm41}. In nature, the universal regime for MHD can be
achieved with $\delta w^\pm \ll v_A$. In numerical
simulations, we can directly solve RMHD equations, which have
precise symmetry already built in. From practical viewpoint, the statistics
from the full MHD simulation with $\delta w^\pm \sim 0.1 v_A$ is virtually
indistinguishable from RMHD statistics and even $\delta w^\pm \sim v_A$ is still fairly similar to the
strong mean field case \citep{BL09a}.

\section{Basic Scalings}
As was shown in a rigorous perturbation study of weak MHD turbulence, it has a tendency of becoming {\it stronger}
on smaller scales \citep{galtier2000}. Indeed, if $k_\|$ is constant and $k_\perp$ is increasing, $\xi=\delta wk_\perp/v_A k_\|$
will increase, due to $\delta w \sim k_\perp^{-1/2}$ in this regime. This will naturally lead to strong turbulence, where
$\xi$ will stuck around unity due to two competing processes: 1) increasing interaction by perpendicular cascade
and 2) decrease of interaction due to the uncertainty relation $\tau_{\rm casc}\omega>1$, where $\tau_{\rm casc}$
is a cascading timescale. Therefore, MHD turbulence will be always marginally strong in the inertial range, which
means that cascading timescale is associated with dynamical
timescale $\tau_{\rm casc}\sim\tau_{\rm dyn}=1/\delta w k_\perp$ \citep{GS95}. 
In this case, assuming that energy transfer is local
in scale and, therefore, depend only
on perturbations amplitude on each scale, we can write Kolmogorov-type phenomenology as

\begin{equation}
\epsilon^+=\frac{(\delta w^+_\lambda)^2\delta w^-_\lambda}{\lambda},
\ \ \ \ \epsilon^-=\frac{(\delta w^-_\lambda)^2\delta w^+_\lambda}{\lambda},
\label{fluxes}
\end{equation}

where $\epsilon^\pm$ is an energy flux of each of the Els\"asser variables and
$\delta w^\pm_\lambda$ is a characteristic perturbation amplitude on a scale
$\lambda$. Such an amplitude can be obtained by Fourier filtering with a
dyadic filter in k-space, see, e.g., \citet{B12a}.

Since we consider so-called balanced case with both $w$'s having the same statistical
properties and energy fluxes, one of these equations is sufficient. This will result
in a $\delta w \sim \lambda^{1/3}$, where $\lambda$ is a perpendicular scale,
or, in terms of energy spectrum $E(k)$,
 
\begin{equation}
E(k)=C_K \epsilon^{2/3} k^{-5/3},\label{spec53}
\end{equation}

where $C_K$ is known as Kolmogorov constant. We will be interested in Kolmogorov constant
for MHD turbulence. This scaling is supposed to work until dissipation effects kick in.
In our further numerical argumentation dissipation scale will play a big role, but
not from a physical, but rather from a formal point of view. We will introduce an idealized
scalar dissipation term in a RHS of Eq.~\ref{mhd2} as $-\nu_n(-\nabla^2)^{n/2}{\bf w^\pm}$,
where $n$ is an order of viscosity and $n=2$ correspond to normal Newtonian viscosity,
while for $n>2$ it is called hyperviscosity. The dissipation scale for this Goldreich-Sridhar
model is the same as the one for Kolmogorov model, i.e.
$\eta=(\nu_n^3/\epsilon)^{1/(3n-2)}$. This is a unique combination of $\nu_n$ and $\epsilon$
that has units of length. Note that Reynolds number, estimated as $v L/\nu_2$, where $L$ is an
outer scale of turbulence, is around $(L/\eta)^{4/3}$.

Furthermore, the perturbations of $w$ will be strongly anisotropic
and this anisotropy can be calculated from the critical balance condition $\xi \approx 1$, so that
$k_\| \sim k_\perp^{2/3}$. Interestingly enough this could be obtained directly from units and
the symmetry of RMHD equations from above. Indeed, in the RMHD limit,
$k_\|$ or $1/\Lambda$ must be in a product with $v_A$, since only the product enters the original
RMHD equations. We already assumed above that turbulence is local and each scale of turbulence
has no knowledge of other scales, but only the local dissipation rate $\epsilon$. In this case
the only dimensionally correct combination for the parallel scale $\Lambda$, corresponding
to perpendicular scale $\lambda$ is 

\begin{equation}
\Lambda=C_A v_A \lambda^{2/3} \epsilon^{-1/3}\label{anis},
\end{equation}

where we introduced a dimensionless ``anisotropy constant'' $C_A$.
Equations ~\ref{spec53} and ~\ref{anis} roughly describe the spectrum
and anisotropy of MHD turbulence.
Note, that Goldreich-Sridhar's $-5/3$ is a basic scaling that should be
corrected for intermittency. This correction is negative due to
structure function power-law exponents being a concave function of their order \citep[see, e.g.,][]{frisch1995}
and is expected to be small in three-dimensional case. This correction for hydrodynamic
turbulence is around $-0.03$. Such a small deviation should be irrelevant in the context
of debate between $-5/3$ and $-3/2$, which differ by about $0.17$.

An alternative model was proposed by \citet{boldyrev2005, boldyrev2006} who suggested
that these scalings are modified by a scale dependent factor that decreases the strength
of the interaction, so that RHS of the Equation~\ref{fluxes} is effectively multiplied
by a factor of $(l/L)^{1/4}$, where $L$ is an
outer scale. We will discuss this hypothesis further in \S 8.
In this case the spectrum will be expressed
as $E(k)=C_{K2} \epsilon^{2/3} k^{-3/2}L^{1/6}$. Note that this spectrum
is the only dimensionally correct spectrum with $k^{-3/2}$ scaling, which
does not contain dissipation scale $\eta$. The absence of $L/\eta$, is due to so-called zeroth
law of turbulence which states that the amplitude at the outer scale should not depend on the viscosity.
This law follows from the locality of energy transfer has been know empirically to hold very well.
The dissipation scale of Boldyrev model is different from that of the Goldreich-Sridhar
model and can be expressed as
$\eta'=(\nu_n^3/\epsilon)^{1/(3n-1.5)}L^{0.5/(3n-1.5)}$.  

\begin{table}
\begin{center}
\caption{Three-dimensional RMHD simulations}
  \begin{tabular*}{1.00\columnwidth}{@{\extracolsep{\fill}}c c c c c c}
    \hline\hline
%Run  & Type & $n_x\times n_y \times n_z$ & Dissipation & $\langle\epsilon\rangle$ & $\eta(\times 10^{-3})$ & $L$ &   $L/\eta$ \\
Run  & $n_x\cdot n_y \cdot n_z$ & Dissipation & $\langle\epsilon\rangle$ &  $L/\eta$ \\

   \hline

R1 & $256\cdot 768^2$ & $-6.82\cdot10^{-14}k^6$   & 0.073 &   200 \\
R2 & $512\cdot 1536^2$ & $-1.51\cdot10^{-15}k^6$  & 0.073  &  400 \\
R3 & $1024\cdot 3072^2$ & $-3.33\cdot10^{-17}k^6$ & 0.073 &  800\\
\hline
R4 & $768^3$ & $-6.82\cdot10^{-14}k^6$            & 0.073 &   200 \\
R5 & $1536^3$ & $-1.51\cdot10^{-15}k^6$            & 0.073 &   400 \\
\hline
R6 & $384\cdot 1024^2$ & $-1.70\cdot10^{-4}k^2$  & 0.081  & 280 \\
R7 & $768\cdot 2048^2$ & $-6.73\cdot10^{-5}k^2$ & 0.081 & 560 \\
\hline
R8 & $768^3$ & $-1.26\cdot10^{-4}k^2$            & 0.073 & 350 \\
R9 & $1536^3$ & $-5.00\cdot10^{-5}k^2$            & 0.073 & 700 \\

   \hline

\end{tabular*}
%\caption{Reduced MHD simulations presented here are divided into 4 groups with similar elementary cell
%geometry, dissipation and driving.}
  \label{experiments}
\end{center}
\end{table}

\section{The Scaling Argument}
Turbulence with very long ranges of scales is common in astrophysics.
Three-dimentional numerics, however, are unable to reproduce such ranges
and normally struggles to obtain even a small inertial range.
In this situation we have to use rigorous quantitative arguments
in order to investigate asymptotic scalings.

Suppose we performed several simulations with different Reynolds
numbers. If we believe that turbulence is universal
and the scale separation between forcing scale and dissipation scale
is large enough, the properties of small scales should not depend on
how turbulence was driven. This is because
neither MHD nor hydrodynamic equations explicitly contain any scale,
so simulation with a smaller dissipation scale could be considered,
due to symmetries from \S 2, as a simulation with the same dissipation
scale, but larger driving scale. E.g., the small scale statistics
in a $1024^3$ simulation should look the same as small-scale statistics
in $512^3$, if the physical size of the elementary cell is the same
and the dissipation scale is the same.

This scaling argument requires that the geometry of the elementary cells is
the same and the actual numerical scheme used to solve the equations
is the same. Also, numerical equations should not contain
any scale explicitly, which is usually satisfied. The
scaling argument does not require high precision on the
dissipation scale or a particular form of dissipation, either
explicit or numerical. This is because we only need
the small-scale statistics to be similar in both simulations.
This is achieved, on one hand, by applying the same numerical
procedure, and, on the other hand, by turbulence locality that
ensures that outer scale influence is small.

In practice, the scaling argument also known as a resolution study is
done in the following way: the averaged spectra in two
simulations are expressed in dimensionless units corresponding
to the expected scaling, for example a $E(k)k^{5/3}\epsilon^{-2/3}$
is used for hydrodynamics, and plotted versus dimensionless
wavenumber $k\eta$, where dissipation scale $\eta$ correspond
to the same model, e.g. $\eta=(\nu^3/\epsilon)^{1/4}$ is used
for scalar second order viscosity $\nu$ and Kolmogorov
phenomenology. As long as the spectra are plotted this way and
the scaling is correct, the curves obtained in simulations with
different resolutions has to converge on small scales.
Not only the spectrum, but any other statistical property of
turbulence can be treated this way. This method has been used in hydrodynamics
since long time ago, e.g. by \citet{yeung1997,gotoh2002,kaneda2003}.
For hydrodynamics good convergence on the
dissipation scale has been obtained with rather moderate resolutions,
which implies that hydrodynamic cascade has fairly narrow locality.
When the resolution is increasing, the convergence is supposed to
become better. This is because of the increased separation of scale
between driving and dissipation.
Better convergence means higher precision in discriminating
different scalings, which was demonstrated in \citet{kaneda2003},
which directly measured the intermittency correction to the spectral slope.
The optimum strategy for our MHD study is to perform the largest resolution
simulations possible and do the scaling study as described in this section.

\section{Numerical methods}
We used pseudospectral dealiased code
to solve RMHD equations. Same code was used earlier for RMHD, incompressible
MHD and incompressible hydrodynamic simulations. The RHS of Eq.~\ref{mhd5}
was complemented by an
explicit dissipation term $-\nu_n(-\nabla^2)^{n/2}{\bf w^\pm}$ and forcing
term ${\bf f}$. The code and the choice for numerical resolution, driving,
etc, was described in great detail in our earlier publications \citep{BL09a,
  BL09b, BL10, B11}. Table 1 shows the parameters of the simulations. The
Kolmogorov scale is defined as $\eta=(\nu_n^3/\epsilon)^{1/(3n-2)}$,
the integral scale $L=3\pi/4E\int_0^\infty k^{-1}E(k)\,dk$ (which was
approximately 0.79 for R1-3). Dimensionless ratio $L/\eta$ could serve as a
``length of the spectrum'', although spectrum is actually significantly shorter
for n=2 viscosity and somewhat shorter for n=6 viscosity.

Since we would like to use this paper to illustrate the resolution
study argument we used a variety of resolution, dissipation and
driving schemes. There are four schemes, presented in Table~1,
and used in simulations R1-3, R4-5, R6-7 and R8-9. In some of
the simulations the resolution in the direction parallel
to the mean magnetic field, $n_x$, was
reduced by a factor compared to perpendicular resolution.
This was deemed possible due to an empirically known lack of
energy in the parallel direction in $k$-space and has been
used before \citep[see, e.g.,][]{muller2005}. The R4-5 and R8-9
groups of simulations were fully resolved in parallel direction.
One would expect that roughly the same resolution
will be required in parallel and perpendicular direction
\citep{BL09a}. In all simulation groups time step was strictly
inversely proportional to the resolution, so that we can utilize
the scaling argument.

Driving had a constant energy injection rate for all simulations
except R6-7, which had fully stochastic driving. All simulations
except R8-9 had Els\"asser driving, while R8-9 had velocity driving.
All simulations were well-resolved and R6-7 were overresolved
by a factor of 1.6 in scale (a factor of 2 in Re).
The anisotropy of driving was that of a box, while injection rate
was chosen so that the amplitude was around unity on outer scale,
this roughly corresponds to critical balance on outer scale.
Indeed, as we will show in subsequent section, since anisotropy
constant is smaller than unity, our driving with $\lambda\sim \Lambda\sim 1$
and $\delta w\sim 1$ on outer scale is somewhat over-critical,
so $\Lambda$ decreases after driving scale to satisfy uncertainty
relation (see Fig.~\ref{anisotropy}). This is good for maintaining critical balance over
wide range of scales as it eliminates possibility for weak turbulence.

\begin{figure*}
%\figurenum{1}
\includegraphics[width=1.0\textwidth]{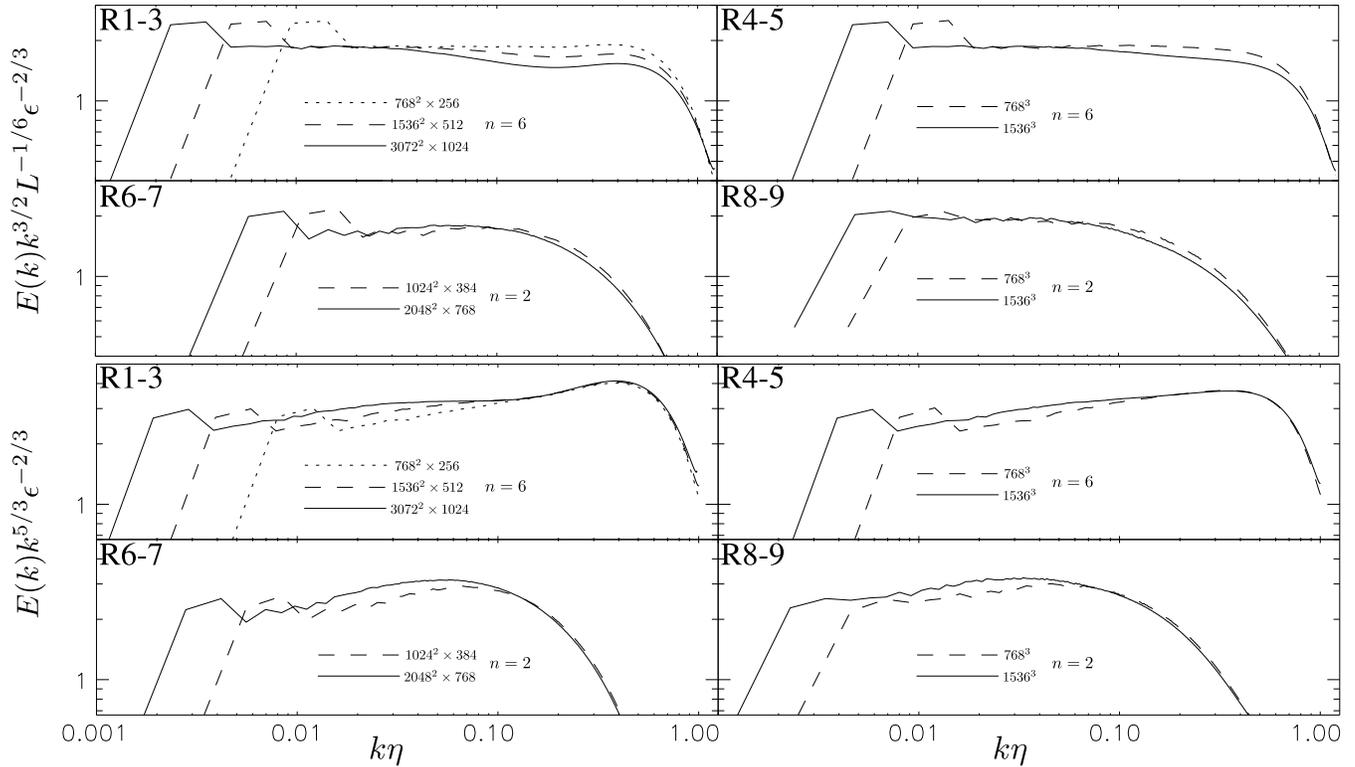}
%\plotone{new_slopes3.eps}
%\plotone{sp_all.eps}
\caption{Numerical convergence of spectra in all simulations. Two upper
rows are used to study convergence assuming Boldyrev model and two
bottom rows -- assuming Goldreich-Sridhar model. Note that definition
of dissipation scale $\eta$ depends on the model, see \S 3, this difference is tiny
in hyperviscous simulations R1-5, but significant in viscous simulations R6-9.
Numerical convergence require that spectra will be similar on small scales, including
the dissipation scale, see, e.g. \citet{gotoh2002}. As we see from the
plots, numerical convergence is absent for Boldyrev model. For Goldreich-Sridhar
model the convergence is reached only at the dissipation scale. Higher-resolution
simulations are required to demonstrate convergence in the inertial range.}
\label{converg}
\end{figure*}

In presenting four groups of simulations, with different
geometries of elementary cell, different dissipation terms
and different driving, our intention is to show
that the scaling argument works irrespective of numerical effects,
but rather relies on scale separation and the assumption of
universal scaling. Simulations R1-3 are the same
as those presented in \citet{B11}.
Largest simulation R3 with $3.1\times10^9$ points used about 0.85 million
CPU hours on TACC Ranger.

\section{Driving and conservation laws}
All of our simulations except R8-9 used Els\"asser driving. The choice of this driving is due
to the formulation of our problem -- we want to obtain asymptotic scalings in the inertial
range of MHD turbulence. This is also the reason we use RMHD, although sometimes full MHD
with strong mean field is used \citep{muller2005, BL09b, BL09a}. In this situation we simulate
a tiny, compared to the outer scale, volume of turbulence. Since incompressible
MHD has two Els\"asser energy cascades
and exchange of energy between these cascades is impossible,
it make sense that we provide energy to these two cascades independently, in fact this is
the only way to simulate imbalanced turbulence in a periodic box \citep{BL09a}.

When driving is concerned, often the issue of integral conservation laws is raised.
In particular, inviscid hydrodynamics has Kelvin's circulation theorem which preserves
circulation along any closed curve, moving with the fluid, as long as the external force
is potential. This also implies that the lines of vorticity are ``frozen`` into the fluid.
Naturally, driving force in most hydrodynamic simulations is not potential
and, as a result, Kelvin's theorem is broken. Although non-potential forces are known
to drive turbulence in nature, this usually is only important on the outer scale of turbulence,
but in the inertial range such forces should be negligible compared to the dynamic pressure.

What is the physical meaning of volumetric force that is typically used in simulations
which try to reproduce inertial range, such as \citet{gotoh2002}? The key to understand this
is to remember that each fluid element is embedded into the surrounding fluid that
exert forces on its boundaries. These forces supply energy that drives turbulence
within this element. Similarly, in MHD fluid an external electromotive force (EMF) exists
on the boundaries of the fluid element and the curl of this EMF enters in RHS of the
induction equation. Unfortunately, simulating fluid element embedded into a larger system
would effectively mean simulating this larger system and is not practical. In practice we use
periodic boxes to mimic statistical homogeneity. Despite driving in such boxes being
restricted to low wavenumbers, the correlation scale is typically smaller than the box
size by a factor of 3-4, which simulates many fluid elements in a box. The driving
around k=2..3, therefore, emulates an action exerted on the boundaries of these fluid
elements. If $k_{\rm max}$ is a maximum wavenumber of
driving, then the change of circulation within a loop of size $l<2\pi/k_{\rm max}$,
associated with forcing, is going to be small, emulating the situation of real
turbulence where the forces exerted on the boundaries of fluid element will actually
preserve circulation along any loop within this element.

\begin{figure}
%\figurenum{1}
\includegraphics[width=0.8\columnwidth]{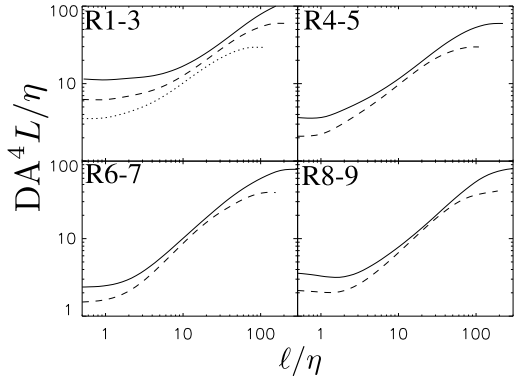}
%\plotone{sp_all.eps}
\caption{Resolution study for ``dynamic alignment``, assuming Boldyrev scaling.
         Both axis are dimensionless,
         solid is higher resolution and dashed is lower resolution.
         Convergence is absent for all simulations. This suggests that $l^{0.25}$
         is not a universal scaling for alignment.}
\label{align_res}
\end{figure}

Similarly, Alfv\'en's theorem is a conservation of the magnetic flux through a given loop
and this implies that the magnetic field lines are frozen into the fluid. It can be broken
by an external EMF, however, if such an EMF is restricted to low wavenumbers in a simulation
with periodic box, this will effectively emulate an external EMF action on the boundaries
of the fluid element. There is a full analogy between Alfv\'en's theorem and Kelvin's theorem
since we expect neither significant external forces nor external EMF at the inertial
range scales. Also, it turns out that there is an analogy between the breakdown of Kelvin's
theorem and Alfv\'en's theorem in turbulent flows \citep{eyink2006, eyink2007}.

To summarize, the driving for a simulation with a strong mean field or RMHD simulation
reproducing inertial range, must have Els\"asser driving, i.e. independent driving of $w^+$ and $w^-$.
This will simulate supply of Els\"asser energies from larger scales. If turbulence
is strong, a pure velocity driving is also possible due to the quick nonlinear
decorrelation of $w^+$ and $w^-$ eddies. Volumetric Els\"asser driving does not honor
Kelvin's or Alfv\'en's theorems, however, by restricting driving to large scales we effectively
emulate the action of external forces which conserve fluxes and circulations.
On the dissipation scales Kelvin's or Alfv\'en's theorems are broken by viscosity
and magnetic diffusivity and in the inertial range they will be broken by turbulence.
Therefore, there is an analogy between forced viscous hydrodynamic simulations
in a periodic box and forced dissipative MHD simulations.

\section{Spectra}

Fig.~\ref{converg} presents a resolution study all simulations. The upper
rows assume Boldyrev scaling, while the bottom rows assume Goldreich-Sridhar scaling.
Reasonable convergence on small scales was achieved only for Goldreich-Sridhar scaling.
The normalized amplitude at the dissipation scale for two upper rows of plots
systematically goes down with resolution, suggesting that $-3/2$ is
not an asymptotic scaling. The flat part of the normalized spectrum on R1-3
plots was fit to obtain Kolmogorov constant of $C_{KA}=3.27\pm 0.07$
which was reported in \citet{B11}. The total Kolmogorov constant
for both Alfv\'en and slow mode in the above paper
was estimated as $C_K=4.2\pm0.2$ for the case of isotropically
driven turbulence with zero mean field, where the energy ratio of slow and Alfven mode
$C_s$ is between 1 and 1.3. This larger
value $C_K=C_{KA}(1+C_s)^{1/3}$ is due to slow mode being passively
advected and not contributing to nonlinearity. The measurement of $C_{KA}$ had relied
on an assumption that the region around $k \eta\approx 0.07$ represent asymptotic
regime. It is possible, though unlikely, that $C_{KA}$ is slightly underestimated
due to this region is somewhat lower than the asymptotic regime due to antibottleneck
effect exacerbated by a reduced parallel resolution. The antibottleneck effect,
however is typically much smaller that the bottleneck effect, so this
correction is probably within the stated errorbars. Next simulation in a R4-5 series
($3072^3$) should make this clear. As far as simulations with normal viscosity R6-9 go,
it seems impossible to approach good inertial range until resolutions of at least
$4096^3$.

\begin{figure}
%\figurenum{1}
%\includegraphics[width=0.9\textwidth]{new_slopes2.eps}
\begin{center}
\includegraphics[width=1.0\columnwidth]{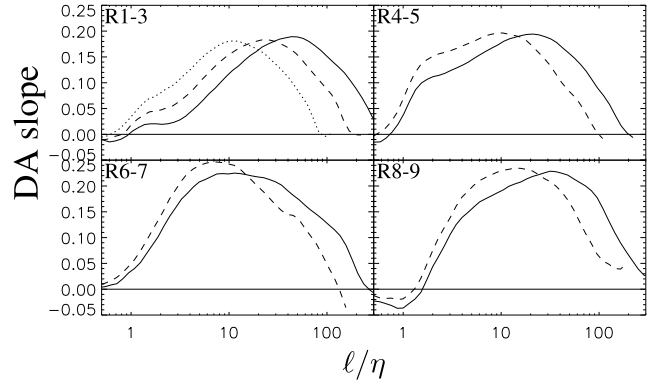}
\end{center}
%\plotone{new_slopes3.eps}
%\plotone{sp_all.eps}
\caption{DA slope, defined as $l/{\rm DA} \partial {\rm DA}/\partial l$,
  solid is higher resolution and dashed is lower resolution. Dynamic alignment slope does not converge and has a tendency
  of becoming smaller in higher-resolution simulations. 
  This may indicate that the asymptotic alignment slope is zero,
  which will correspond to the GS95 model.}
\label{align_sl_da}
\end{figure}

\section{Dynamic alignment}

\citet{boldyrev2005} suggested that ${\bf w}^+$ and ${\bf w}^-$ eddies
are systematically aligned. This was investigated numerically in
in \citet{BL06} and no significant alignment was found for the averaged
angle between ${\bf w}^+$ and ${\bf w}^-$,
$AA=\langle|\delta {\bf w}^+_\lambda\times \delta {\bf w}^-_\lambda|/|\delta
{\bf w}^+_\lambda||\delta {\bf w}^-_\lambda|\rangle$, but when this angle
was weighted with the amplitude $PI=\langle|\delta {\bf
  w}^+_\lambda\times \delta {\bf w}^-_\lambda|\rangle/\langle|\delta {\bf
  w}^+_\lambda||\delta {\bf w}^-_\lambda|\rangle$, some alignment was found.
Then \citet{boldyrev2006} suggested alignment between ${\bf v}$
and ${\bf b}$ and \citep{mason2006} suggested a particular
amplitude-weighted measure, $DA=\langle|\delta {\bf
  v}_\lambda\times \delta {\bf b}_\lambda|\rangle/\langle|\delta {\bf
  v}_\lambda||\delta {\bf b}_\lambda|\rangle$. We note that DA is similar
to PI but contain two effects: alignment and local imbalance. The
latter could be measured with
$IM=\langle |\delta (w^+_\lambda)^2- \delta (w^-_\lambda)^2|\rangle /\langle \delta
(w^+_\lambda)^2+ \delta (w^-_\lambda)^2\rangle$, \citet{BL09b}.

In this section we check the assertion of \citet{boldyrev2005, boldyrev2006}
that alignment depends on scale as $\lambda^{1/4}$, by using DA which is, by
some reason, favored by aforementioned group. We did a resolution study of DA,
assuming suggested scaling, which is presented on Figure~\ref{align_res}.
Convergence was absent in all simulations. Note, that
previous studies that claimed that there is a good correspondence with
Boldyrev model did not perform the scaling study, therefore these
claims are not well substantiated. A result
from a single isolated simulation could be easily contaminated by the effects
of outer scale, since it is not known a-priori how local MHD turbulence is
and what resolution is sufficient to get rid of such effects. On the contrary,
the resolution study offers a systematic approach to this problem.

Fig.~\ref{align_sl_da} shows ``dynamic alignment`` slope for all simulations. Although
there is no convergence as in the previous plot, it is interesting
to note that alignment slope decreases with resolution. This suggests
that most likely the asymptotic state for the alignment slope is zero,
i.e. alignment is scale-independent and Goldreich-Sridhar model is
recovered. Also, alignment from simulations R1-5 seems to indicate
that the maximum of the alignment slope is tied to the outer scale,
therefore alignment is a transitional effect.

\begin{figure}
%\figurenum{1}
%\includegraphics[width=0.9\textwidth]{new_slopes2.eps}
\begin{center}
\includegraphics[width=1.0\columnwidth]{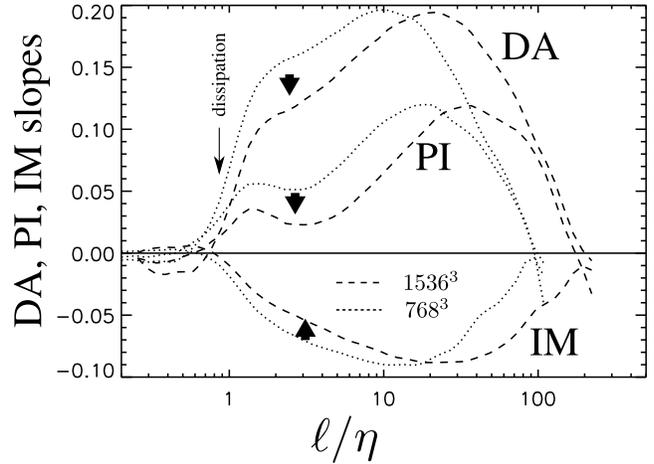}
\end{center}
%\plotone{new_slopes3.eps}
%\plotone{sp_all.eps}
\caption{Slopes of several alignment measures vs scale in R4-5 (for definitions see the text).
  Each measure follows its own scaling, however there are indications that they are all
  tied to the outer scale, due to the maximum of alignment being a fraction
  of the outer scale, which is an indication that their scale-dependency is of transient nature.}
\label{align_sl_all}
\end{figure}

In our earlier studies \citep{BL06,BL09b} we measured several types
of alignment and found no evidence that all alignment measures
follow the same scaling, see, e.g., Fig.~\ref{align_sl_all}.
As one alignment measure, PI, has been already known to be scale-dependent \citep{BL06} prior
to DA, it appears that a particular measure of the alignment in \citet{mason2006} was
been picked for being most scale-dependent and no thorough explanation
was given why it was preferred. Futhermore, it was claimed that DA has the asymptotic
scaling of $\sim (l/L)^{1/4}$ and at the same time it is an interaction weakening factor
that will result in a $-3/2$ spectrum. This is unlikely, since the interaction
weakening factor will appear from a third-order structure function, while DA is a
correction factor based on a ratio of second-order structure functions. Intermittency
corrections are supposed to be small, however. More importantly, a physical justification
of DA and its preference over other measures, such as PI or IM, that differ from DA
quite significantly, was lacking.

We are not aware of any convincing physical argumentation explaining why
alignment should necessarily be a power-law of scale. \citet{boldyrev2006}
argues that alignment will tend to increase, but
will be bounded by field wandering, i.e. the alignment on each scale will be
created independently of other scales and will be proportional to the relative
perturbation amplitude $\delta B/B$. But this violates two-parametric
symmetry of RMHD equations mentioned above, which suggests that
field wandering can not destroy alignment or imbalance. Indeed, a perfectly
aligned state, e.g., with $\delta {\bf w}^-=0$ is a precise solution of
MHD equations and it is not destroyed by its own field wandering.
The alignment measured in simulations of strong MHD
turbulence with different values of $\delta B_L/B_0$ showed very little or
no dependence on this parameter \citep{BL09b}. 

Why alignment measures are scale-dependent quantities over about
one order of magnitude in scale is an interesting question.
Most plausible explanation is that because MHD turbulence is much
less local than hydro turbulence \citep{BL09b, BL10, B11} and
because driving does not mimic the properties of the inertial
range, the transition to asymptotic statistics is very wide
and many quantities appear as scale-dependent, while they are
simply adjusting to the asymptotic regime. 

The contribution to energy flux from different $k$
wavebands is important to understand, since most cascade models
assume locality, or rather to say the very term ''cascade'' assumes
locality. An analytical upper bound on locality suggests
that the width of the energy transfer window can scale as $C_K^{9/4}$
\citep{B12a}, however, in practice turbulence
can be more local. The observation of \citet{BL09b} that MHD simulations
normally lack bottleneck effect, even with high-order dissipation,
while hydrodynamic simulations always have bottleneck, which is
especially dramatic with high-order dissipation, is consistent with
above conjecture on locality, since bottleneck effect relies
on locality of energy transfer.
As locality constraint depends on the efficiency
of the energy transfer, so that the efficient energy transfer must be local,
while inefficient one could be nonlocal \citep{BL10,B11,B12a}.
As we observe larger $C_K$ in MHD turbulence compared to hydrodynamic
turbulence, the former could be less local than the latter, which
is consistent with our earlier findings.

\begin{figure}
%\figurenum{1}
%\includegraphics[width=0.9\textwidth]{new_slopes2.eps}
\begin{center}
\includegraphics[width=1.0\columnwidth]{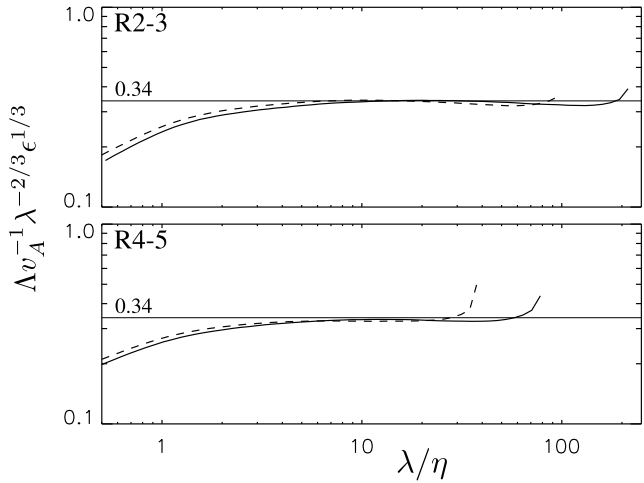}
\end{center}
%\plotone{new_slopes3.eps}
%\plotone{sp_all.eps}
\caption{The scaling study for anisotropy shows moderately good
convergence to a universal anisotropy $\Lambda=C_A v_A \lambda^{2/3} \epsilon^{-1/3}$ with
anisotropy constant $C_A$ of around 0.34.}
\label{anisotropy}
\end{figure}

\section{Anisotropy}
In Section 3 we suggested that anisotropy
should be universal in the inertial range
and expressed as $\Lambda=C_A v_A \lambda^{2/3} \epsilon^{-1/3}$,
where $C_A$ is an anisotropy constant to be determined from the numerical
experiment or observation. Note, that both Alfv\'enic and slow modes
should have the same anisotropy. This is because they have the same
ratio of propagation to nonlinear timescales. Figure~\ref{anisotropy} shows
anisotropy for the two best resolved groups R1-3 and R4-5. We used
a model independent method of minimum parallel structure function,
described in detail in \citet{BL09a}. Alternative definitions
of local mean field give comparable results, as long as they are reasonable.
The convergence of anisotropy curves at the dissipation scale is not as good
as convergence of spectra in R1-5 on the dissipation scale. This is due
to this measure being calculated from structure functions which are
less local in scale than 3D spectra, i.e. disipation scale is still being
somewhat affected by the transition to asympotic regime.
From R1-3 we obtain $C_A=0.34$. Note, that the conventional definition of
critical balance involve the amplitude, rather than $(\epsilon\lambda)^{1/3}$,
so the constant in this classical formulation will be $C_A C_K^{1/2}\approx 0.63$,
which is closer to unity. Together with energy spectrum this is a full
description of universal axisymmetric two-dimensional spectrum of MHD
turbulence in the inertial range.

\section{Summary and Discussion}

In this paper we argued that the properties of Alfven and slow components
of MHD turbulence in the inertial range will be determined only by the Alfven speed
$v_A$, dissipation rate $\epsilon$ and the scale of interest $\lambda$. The energy spectrum
and anisotropy of Alfven mode will be expressed as

$$
E(k)=C_K \epsilon^{2/3} k^{-5/3},
$$
$$
\Lambda/\lambda=C_A v_A (\lambda \epsilon)^{-1/3},
$$

with $C_K=3.3$ and $C_A=0.34$. If the slow mode is present, its anisotropy will
be the same, and it will contribute to both energy and dissipation rate. Assuming
the ratio of slow to Alfven energies between 1 and 1.3, the latter was observed in
statistically isotropic high resolution MHD simulation with no mean field, we can
use $C_K=4.2$ for the total energy spectrum \citep{B11}.

Anisotropy of MHD turbulence is an important property that affects such processes
as interaction with cosmic rays, see, e.g., \citet{yan2002}. Since cosmic ray
pressure in our Galaxy is of the same order as dynamic pressure, their importance
should not be underestimated. Another process affected is the three-dimentional
turbulent reconnection, see, e.g., \citet{lazarian1999}.

Previous measurements of the energy slope relied on the
highest-resolution simulation and fitted the slope in the fixed $k$-range
close to the driving scale, typically between $k=5$ and $k=20$. We argue that such
a fit is unphysical unless a numerical convergence has been demonstrated.
We can plot the spectrum vs dimensionless $k\eta$ and if we clearly
see a converged dissipation range and a bottleneck range, we can assume that
larger scales, in terms of $k \eta$ represent inertial range.
In fitting fixed $k$-range at low $k$ we will never
get rid of the influence of the driving scale. In fitting a fixed $k \eta$
range, the effects of the driving will diminish with increasing resolution.

Since we still have trouble transitioning into the inertial range in 
large mean field simulations, for now it is impossible to 
demonstrate inertial range in statistically isotropic simulations
similar to once presented in \citet{muller2005}. This is because
we do not expect a universal power-law scaling in transAlfv\'enic
regime, due to the absence of appropriate symmetries and the
transitioning to subAlfv\'enic regime, where such scaling is possible,
will require some extra scale separation.
These two transitions require numerical resolution that is
even higher than the highest resolution presented in this paper and for now
seem computationally impossible.

Full compressible MHD equations contain extra degrees of freedom, which, in a weakly compressible case, entails the
additional cascade of the fast MHD mode, possibly of weak nature. Supersonic simulations with
moderate Mach numbers \citep{cho2003c} show that Alfv\'enic cascade is pretty resilient and is not much affected
by compressible motions. The models of the "universal" supersonic turbulence covering supersonic
large scales and effectively subsonic small scales are based mainly on simulations with limited resolution
and unlikely to hold true. This is further reinforced by the results of this paper which demonstrated that even
a much simpler case of sub-Alfv\'enic turbulence require fairly high resolutions to obtain an asymptotic scaling.  

In this paper we treated so called balanced case with $\delta w^+\approx \delta w^-$.
A more general imbalanced case has been discussed in \citet{BL08, BL09a, BL10}, see also
references therein.

\subsection{Results of \citet{grappin2010}}
\citet{grappin2010} observed that MHD turbulence have scale-independent
anisotropy, in contrast with the Goldreich-Sridhar model and
our own results. We believe that this measurement is correct,
in fact, similar 2D spectra has been reported in our study \citet{BL09a}.
However, these are 2D spectra obtained with respect to the
global mean magnetic field. A trivial exercise \citep{CLV02a,BL09b} show that measuring
anisotropy with respect to the global field will destroy scale-dependent
anisotropy, even in the case of very strong field. Indeed, even if we have
$\delta B_L/B_0 \ll 1$, the anisotropy in the global frame will
be limited from above by $B_0/\delta B_L$, due to field wandering, while
the Goldreich-Sridhar anisotropy on the scale $l$ will be
much higher, $\sim B_0/\delta B_l$, by a factor of $B_L/B_l$.
In particular, each individual volume on scale $l$ will have
anisotropy with respect to the mean field averaged on the same
volume, however, the direction of this field will deviate
from the global mean field statistically, with RMS deviation
of around $\delta B_L/B_0$. This directional deviations will
easily destroy anisotropy $\sim B_0/\delta B_l$ if we average
spectra measured with respect to $B_0$.
Therefore, one must measure anisotropy with respect to local mean
field, which was realized in \citet{Cho2000}. From
this argument, we see that the local mean field must be averaged
on scales smaller or equal to $l$, a scale at which spectrum or structure
function is measured and $l$ itself is a preferred averaging scale.
Also, it is anisotropy with respect to the local mean field, which 
is important for cosmic ray dynamics \citep{yan2002, yan2004}.

\subsection{Results of \citet{mason2011}}
\citet{mason2011} conducted a series of low-resolution simulations ($256^3-512^3$) studying
the effect on dynamic alignment by numerics on viscous scale,
e.g., by changing resolution with constant $Re$, or changing elementary cell geometry,
such as the ratio of parallel to perpendicular resolution. Their conclusion was
that alignment is, indeed, somewhat influenced by numerics.

As we explained previously, numerical effects on viscous scale are
irrelevant for scaling study, if latter was conducted properly. Therefore it was quite
puzzling that \citet{mason2011} claimed that their results invalidate \citet{B11},
which did a proper scaling study with high resolution simulations. Furthermore,
it is also puzzling that \citet{mason2011} claimed an ``extended scaling
law'' for alignment (previously mentioned $\lambda^{1/4}$), despite the fact that
they did not even attempt a proper scaling study for this quantity and did not
demonstrate convergence. 

Although studying obscure effects on viscous scales is pretty common in modern
numerical studies of hydrodynamic turbulence, we will argue that such studies
in MHD are of limited value for astrophysics, since dissipation mechanism
of astrophysical plasma is, typically, not a scalar second-order diffusion.

On the contrary, an asymptotic scalings of the inertial range of MHD turbulence
is of high value. However, such scalings could only be demonstrated by a proper
rigorous scaling study in the spirit of \citet{yeung1997, gotoh2002}, and not
by arbitrarily assigning ``inertial range'' to a fixed range in wavenumbers.

\section*{Acknowledgments}
%\begin{acknowledgments}
%LA-UR 10-07570.

I am grateful to Miriam Forman, Rob Wicks, Tim Horbury, Giga Gogoberidze, Alex Schekochihin and Steve Cowley
for illuminating discussions. The author was supported by LANL Director's Fellowship and Humboldt Fellowship.
Computations were performed on TACC Ranger through
NSF TeraGrid allocation TG-AST080005N.

\par\ \ \ 

\par

\def\apj{{\rm ApJ}}           
\def\apjl{{\rm ApJ }}          
\def\apjs{{\rm ApJ }}          
\def\grl{{\rm GRL }}
\def\aap{{\rm A\&A } }
\def\mnras{{\rm MNRAS } }
\def\physrep{{\rm Phys. Rep. } }               
\def\prl{{\rm Phys. Rev. Lett.}} 
\def\pre{{\rm Phys. Rev. E}} 
\bibliography{all}

\end{document}